\documentclass[aps,prl,twocolumn,groupedaddress]{revtex4}

\usepackage{graphicx}

\begin{document}
%\draft
%\preprint{{\bf ETH-TH/98-??}}

\title{Commensurate-incommensurate transition of cold atoms in an
optical lattice}

\author{H.P.\ B\"uchler}
\author{G.\ Blatter}
\author{W.\ Zwerger}
\altaffiliation{ Present and permanent address: Sektion Physik,
Universit\"at M\"unchen, Theresienstr.\ 37, D-80333 M\"unchen,
Germany.}
\affiliation{Theoretische Physik, ETH-H\"onggerberg, CH-8093 Z\"urich,
Switzerland}

\date{\today}

\begin{abstract}
  An atomic gas subject to a commensurate periodic
  potential generated by an optical lattice undergoes a
  superfluid--Mott insulator transition. Confining a strongly
  interacting gas to one dimension generates an instability
  where an arbitrary weak potential is sufficient to pin the
  atoms into the Mott state; here, we derive the corresponding
  phase diagram. The commensurate pinned state may be detected
  via its finite excitation gap and the Bragg peaks in the static
  structure factor.
\end{abstract}

%\pacs{03.75.Fi, 05.30.Jp, 32.80.Pj}

\maketitle

Atomic gases are developing into the ultimately tunable laboratory
system allowing to study complex quantum phenomena
\cite{anglin02}. Recently, subjecting an atomic Bose-Einstein
condensate to an optical lattice, Greiner {\it et al.}
\cite{greiner02} have succeeded in tuning the system through a
quantum phase transition separating a superfluid (S) from a Mott
insulating (MI) phase. This 3D bulk transition involves weakly
interacting bosons and is well understood within the Bose-Hubbard
description \cite{fisher89,jaksch98}: the system turns insulating when the
on-site interaction energy $U$ becomes of the order of the hopping
energy $J$. This strong coupling transition is a result of
quenching the kinetic energy by a strong lattice potential.
Amazingly, by confining the atomic gas to one dimension, the
strong coupling limit can be reached {\it without} the optical
lattice: in 1D, the ratio $\gamma$ between the interaction- and
kinetic energies per particle scales inversely with the density 
$n$ and thus it is the {\it low}-density limit which is
interacting strongly (Tonks gas) \cite{petrov00}. A new
instability then appears in the strongly interacting 1D quantum
gas: the superfluid groundstate in the homogeneous system turns
insulating in the presence of an arbitrarily weak commensurate
optical lattice and the S--MI transition changes to a transition of
the incommensurate--commensurate type. In this letter, we
analyze this new instability and derive the complete phase diagram
for the S--MI transition in the limits of both weakly and strongly
interacting gases. Remarkably, this goal can be achieved by a
mapping to two classic problems, the Bose-Hubbard model
(introduced in Ref.\ \onlinecite{jaksch98})
and the sine-Gordon problem describing the weakly- and strongly
interacting limits of the atomic gas, respectively. Below, we
first summarize the main results providing us with the phase
diagram shown in Fig.\ 1; we then proceed with a detailed analysis
of the strongly interacting Bose gas subject to a weak optical
potential, leading us to a proper understanding of the new
instability.

\begin{figure}[hbtp] \label{Phasedia}
\includegraphics[scale=0.32]{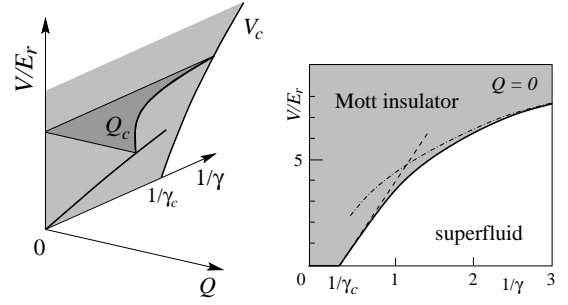}
  \caption{Left: Schematic phase diagram illustrating the 
  superfluid and Mott insulating phases versus parameters
  $\gamma$ (interaction), $V$ (optical potential), and $Q$
  (commensuration).  
  Right: Critical amplitude $V_{c}$ versus 
  interaction $1/\gamma$ for the commensurate
  situation with $Q=0$. Below $1/\gamma_{c}$,
  an arbitrary weak potential $V$ drives the superfluid into
  the pinned insulating state. The dashed line denotes the
  asymptotic behavior near the critical point $1/\gamma_{c}$ as
  determined from the sine-Gordon model,
  while the dashed-dotted line derives from the Bose-Hubbard
  criterion $U/J|_{\rm\scriptscriptstyle S-MI} \approx 3.84$.}
\end{figure}

A weakly interacting atomic gas subject to an optical lattice is
well described by the Bose-Hubbard model, which starts from a
tight-binding model and takes the interaction between bosons into
account perturbatively; the hopping amplitude $J(V)$ and the
on-site interaction energy $U(V,\gamma)$ follow from the
underlying parameters of the atomic gas, the dimensionless
interaction parameter $\gamma$ and the amplitude $V$ of the
optical lattice. The phase diagram of the Bose-Hubbard model is
well known \cite{fisher89} and involves insulating Mott-lobes
embedded in a superfluid phase. In 3D, the mean-field analysis for
densities commensurate with the lattice provides the critical
parameter $U/J|_{\rm\scriptscriptstyle S-MI} \approx 5.8\, z$, in
good agreement with the experimental findings of Greiner {\it et
al.} \cite{greiner02} (here, $z$ denotes the number of nearest
neighbors). Going to 1D, fluctuations become important and
appreciably modify the mean-field result: numerical simulations
\cite{kuhner98,rapsch99} place the transition at the critical
value $U/J|_{\rm\scriptscriptstyle S-MI} = 2C \approx 3.84$. This result
is easily transformed into the $\gamma$--$V$ phase diagram of the
weakly interacting atomic gas, once the relations $J(V)$ and
$U(V,\gamma)$ to the Mott-Hubbard parameters are known; combining
$J \propto \exp[-2(V/E_r)^{1/2}]$ as derived from a
WKB-calculation, with the on-site repulsion $U \propto \gamma$,
the transition line $V_c(\gamma)$ (see Fig.\ 1) then derives from
the implicit equation
\begin{equation}
   4V/E_{r} = \ln^{2}\bigl[ 4\sqrt{2} \pi C
   \left(V/E_{r} \right)^{1/2}/\gamma\bigr].
\end{equation}
Here, $E_{r}= \hbar^{2}k^{2}/2m$ is the recoil energy with $m$ the
boson mass and $k=2\pi/\lambda$ the wave vector of the light
generating the optical lattice $V(x) = V \sin^2(kx)$. 
The dimensionless interac\-tion strength
$\gamma$ is defined via $\gamma= m g/\hbar^{2}n$, with $n$ the
density and $g$ the strength of the $\delta$-function interaction
potential; $g$ is related to the 3D scattering length $a_{s}$ and
the transverse confining frequency $\omega_{\perp}$ via $g = 2
\hbar \omega_{\perp} a_{s}$ \cite{olshanii98}.

Increasing the interaction strength $\gamma$, the critical
amplitude $V_c$ of the optical lattice triggering the S--MI
transition decreases, see Fig.\ 1; the description of the atom gas
in terms of the Bose-Hubbard model breaks down and we have to look
for a new starting point. For a weak optical potential, a natural 
choice is the 1D Bose gas with $\delta$-function interaction,
which resides in the strong coupling regime at small densities 
$n$ \cite{petrov00}; the presence of the optical lattice is 
taken into account perturbatively. The homogeneous 1D Bose gas
with $\delta$-function interaction has been solved exactly by Lieb
and Liniger \cite{lieb63}; the corresponding low-energy
physics is properly described in terms of a Luttinger liquid with
a parameter $K(\gamma)$ derived from the exact solution, see
below. Adding the optical potential $\propto V$, we arrive at the
sine-Gordon model; the critical value $K(\gamma,V) = 2(1+V/4E_r)$
separating the Mott insulating phase from the superfluid one
determines the phase line $V_c/E_{r} = (\gamma^{-1}-
\gamma_{c}^{-1})/5.5$, with the critical value $\gamma_{c}\approx
3.5$. Combining the results of the Bose-Hubbard- and sine-Gordon
models we can complete the phase diagram for the commensurable
situation as shown in Fig.\ 1: most remarkable is the appearance
of a critical interaction strength $\gamma_c$ above which an
arbitrary weak optical lattice is able to pin the system into a
Mott insu\-lator state. The presence of this instability is due to
the closeness of the dilute 1D Bose liquid to
Wigner-crystallization. Tuning the system away from
commensurability with $Q \equiv 2\pi(n-2/\lambda) \neq 0$ the Mott
insulator survives up to a critical misfit $Q_c(V,\gamma)$, see
Fig.~1. The corresponding physics is similar to that of the
com\-men\-su\-rate-in\-com\-mensurate transition of adsorbates on
a per\-iodic substrate as studied by Pokrovsky and Talapov
\cite{pokrovsky79}.

For a qualitative understanding of the transition in the strongly
interacting 1D Bose gas, it is useful to consider the limit
$\gamma \gg 1$, where the behavior is essentially that of an ideal
Fermi gas \cite{girardeau60}. A weak periodic potential
$V(x)=V\sin^{2}(k x)$, with a lattice constant such that an
integer number $i=1,2,\ldots$ of particles will fit into one unit
cell, gives rise to a single-particle band structure in which the
$i$ lowest bands are completely filled. The ground state for
noninteracting Fermions is then a trivial band insulator,
separated from excited states by an energy gap which scales like
$(V/E_r)^i$ in the limit $V \ll E_{r}$. Similar to the Mott phase
in the Bose-Hubbard model, the insulating state has a fixed
integer density, commensurate with the lattice. It remains locked
in a finite regime of the chemical potential, characterizing an
incompressible state \cite{fisher89}. Clearly, for a weak periodic
potential, the lowest energy gap with $i=1$ is much larger than
the higher-order ones. In the following, we will thus confine
ourselves to studying the commensurate-incommensurate transition
near integer filling $i=1$, where the commensurate phase has
maximal stability.

For a quantitative theoretical analysis, it is convenient to use
Haldane's description of the interacting 1D Bose gas in terms of
its long wave length density oscillations \cite{haldane81}.
Introducing the two fields $\phi(x)$ and $\theta(x)$ describing
phase and density fluctuations, the Hamiltonian of the homogeneous
gas (without longitudinal confining trap and periodic potential)
is a quadratic form involving kinetic and interaction energies,
\begin{equation}
   H_{0} = \frac{\hbar}{2\pi}
   \int dx \left[ v_{J} (\partial_{x}\phi)^{2}
   + v_{N} (\partial_{x}\theta- \pi n)^{2}\right].
   \label{hamiltonian}
\end{equation}
Here, $v_{J}=\pi \hbar n/m$ is the analog of the bare `Fermi'
velocity at given average density $n$ while $v_{N}=\partial_{n}
\mu/\pi \hbar$ is determined by the inverse compressibility,
giving a sound velocity $v_{s}=\sqrt{v_{J} v_{N}}$ consistent with
the standard thermodynamic relation  $mv_{s}^{2} = n \partial_{n}
\mu$. For short-range repulsive interactions, the dimensionless
ratio $K=\beta^{2}/4\pi= v_{J}/v_{s}$ is larger than one,
approaching $\infty$ in the noninteracting case and unity in the
hard core limit \cite{haldane81}. In the idealized model with
$\delta$-function interactions, the exact solution by Lieb and
Liniger shows that $K$ is a monotonically decreasing function of
the ratio $\gamma= m g/\hbar^{2}n$ between the interaction and
kinetic energies; the limiting behavior for small values of
$\gamma$
\begin{equation}
   K(\gamma \rightarrow 0)
   = \pi [\gamma-(1/2 \pi) \gamma^{3/2}]^{-1/2}
\label{kgamma}
\end{equation}
follows from the Bogoliubov approximation in 1D. Surprisingly,
this result remains quantitatively correct for $\gamma$ values up
to $10$ \cite{lieb63}. At large $\gamma>10$, the asymptotic
behavior is $K(\gamma\rightarrow \infty)=(1+2/\gamma)^2$.

For our subsequent analysis, it is convenient to introduce the
conjugate fields $\Theta(x)= (2/\beta)\left[\theta(x) - \pi n
x\right]$ and $\Pi(x)=- \hbar (\beta/2 \pi) \partial_{x} \phi(x)$
obeying the commutation relation $\left[ \Theta(x), \Pi(x')\right]
= i \hbar \delta(x-x')$. Using these fields,
\begin{equation}
   H_{0} = \frac{\hbar v_{s}}{2}
   \int dx \left[ \left(\Pi/\hbar\right)^{2}
   + \left(\partial_{x} \Theta \right)^{2}\right]
   \label{string}
\end{equation}
takes the form of the Hamiltonian for a 1D harmonic string with a
linear spectrum $\omega = v_{s} q$ describing the long wave length
density modulations of the interacting Bose gas. The assumption of
a linear spectrum inherent in the simple form (\ref{string}) is
valid only below a momentum cutoff $1/a \sim \pi n$
\cite{lieb63}; the choice of the length scale $a$ fixes the
energy scale of $H_0$. The Boson density operator $n(x)$ is
related to the field $\Theta(x)$ by \cite{haldane81}
\begin{equation}
   n(x) = \biggl[n+\frac{\beta}{2 \pi} \partial_{x}\Theta\biggr]
   \biggl[1 + 2 \sum_{l =1}^{\infty}
   \cos\biggl( \frac{l\beta \Theta}{2}+l \pi n x\biggr)\biggr];
   \label{densityoperator}
\end{equation}
the last factor accounts for the discrete nature of the particles.
Adding an external periodic potential with amplitude $V/2$ and
period $\lambda/2$ gives rise to the perturbation
\begin{equation}
  H_{\rm \scriptscriptstyle V}
  =\frac{V}{2} \int dx\: n(x) \cos \frac{4 \pi x}{\lambda}.
\end{equation}
As noted already by Haldane, insertion of the Fourier expansion
(\ref{densityoperator}) generates terms of the type appearing in
the quantum (1+1)-dimensional sine-Gordon theory
\cite{coleman75,gogolin98}. Close to commensurability the dominant
term arising from the lowest harmonic in (\ref{densityoperator})
has the conventional sine-Gordon form \cite{fisher89}
\begin{equation}
   H_{\rm \scriptscriptstyle V} = \frac{V\:n}{2}\,
   \int dx\,\cos \left[ \beta \Theta + Q x\right]
\end{equation}
with coupling parameter $\beta = 2 (\pi K)^{1/2}$ and a twist
$Q=2\pi(n-2/\lambda)$. The strength of the nonlinear $\cos \beta
\Theta$- perturbation is conveniently expressed through the
dimensionless parameter $u=\pi a^{2} n V/2\hbar v_{s}$ which
naturally involves the cutoff parameter $a$ \cite{kehrein99_01}. 
The twist $Q$
vanishes at commensurability; away from commensurability the
finite twist $Q$ acts as a chemical potential for excitations and
is preferably incorporated into the free Hamiltonian
(\ref{string}) via the replacement $\partial_{x} \Theta
\rightarrow \partial_{x} \Theta-Q/\beta$.

At fixed potential $V$, the quantum sine-Gordon model describes
the competition between the preferred average inter-particle
distance at given density due to the repulsive interaction and the
period imposed by the external potential. A perturbative
calculation (see Ref.~\onlinecite{gogolin98} for a review) tells
that for $\beta^{2}/4 \pi =K>2$ a weak periodic potential is
unable to pin the density; hence for $K>K_{c}=2$
($\gamma<\gamma_{c} \approx 3.5$) the ground state remains gapless
and superfluid in the presence of a small-amplitude lattice
potential. In the strong coupling regime $K<2$, however, the atoms
are locked even to a weak lattice, as long as the twist $Q$ is
less than a critical value $Q_{c}$. Beyond that, there is a finite
density of `solitons' (or domain walls for adsorbates on a
periodic substrate, see Ref.\ \onlinecite{pokrovsky79}), which
interpolate between minima of the external potential, relieving
the frustration present at incommensurate densities $Q\neq0$. The
`solitons' behave like relativistic particles with energy $E_{q} =
\hbar v_{s} \sqrt{q^{2}+M^{2}}$ and reestablish the superfluid
response. The `mass' $M$ determines the excitation gap in the Mott
insulating state, which translates into a jump $\Delta \mu$ in the
chemical potential at the commensurate density: given that an
additional/missing atom involves $K$ solitonic excitations with
energy $E_{q=0}=\hbar v_{s} M$ \cite{japaridze84} one obtains
\begin{equation}
  \Delta\mu =\frac{2\pi\hbar^{2}n}{m}\, M;
\end{equation}
furthermore, this mass is also simply related to the critical
twist via $Q_{c} =2 K^{2}M$. The precise numerical value of $M$
depends on the high momentum cutoff $1/a$ via the dimensionless
amplitude $u$ of the lattice potential. The free-fermion limit
$K=1$ fixes this cutoff at $1/a=\pi n$, resulting in the simple form
$u= V/4 E_{r}$; we ignore small corrections arising due to a 
possible modification in the cutoff away from $K=1$. The dependence 
of the mass $M$ on $u$ can be obtained from a recent nonperturbative
renormalization group analysis of the quantum sine-Gordon model by
Kehrein \cite{kehrein99_01}; for small values $V \ll E_{r}$
and $K$ away from $K_c$, the gap in the chemical potential takes
the form
\begin{equation}
   \Delta\mu = 2 E_{r}\,\left[\frac{V}
    {(2-K)4 E_{r}}\right]^{1/(2-K)}.
    \label{mu_K}
\end{equation}
In the limit $K\rightarrow 1$, the size of the gap approaches
$V/2$, in agreement with the above fermionic picture of the Tonks
gas limit, where the appearance of an insulating ground state,
even for a weak periodic potential, is due to the opening of a
single-particle band gap at the Fermi energy. In the practically
accessible regime of large but finite $\gamma>\gamma_{c}$ the gap
depends on $V$ in the more complicated manner as given by
(\ref{mu_K}) and vanishes exponentially as $K$ approaches
$K_{c}=2$ \cite{kehrein99_01}, see Fig.\ 2(a). Similarly,
the density range $n-2/\lambda= \pm Q_{c}/2\pi$ over which the
ground state remains locked approaches zero as $K\rightarrow K_c$.
Finally, for $K>2$, the dependence of the critical interaction
parameter $K_c$ (or $\gamma_c$) on the lattice amplitude $V$
follows easily from the Kosterlitz-Thouless nature of the scaling
flow near $K_{c}=2$: to lowest order in $u$, $K_{c}(u)= 2(1+u)$.
Combining this result with (\ref{kgamma}) it is straightforward to
determine the line $V_c(\gamma)$ separating the gapped insulating
regime from the superfluid at small but finite values of $u$, see
Fig.\ 1.
\begin{figure}[hbtp]
 \includegraphics[scale=0.38]{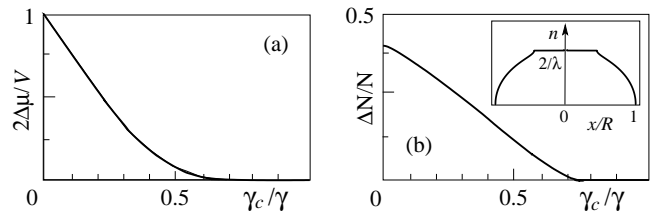}
  \caption{(a) Size of the gap $\Delta \mu$ versus interaction
  strength $\gamma$ for a fixed amplitude $V = E_{r}/2$.
  For $\gamma \rightarrow \infty$ the gap assumes the 
  free fermion limit $V/2$, while it vanishes exponentialy 
  for $\gamma \rightarrow \gamma_{c}$.
  (b) Fraction of atoms in the Mott insulating phase. The inset
  shows the density distribution $n(x)$ with the Mott phase
  characterized by a locked commensurate density in the trap
  center, surrounded by a superfluid region.
\label{excitationgap}}
\end{figure}

In order to analyze the consequences of the
commensu\-rate-incommensurate transition for cold atoms in a trap,
we consider a 1D Bose gas in the presence of a weak longitudinal
confining potential $V(x) = m \omega^{2}x^{2}/2$. Provided the
associated oscillator length $l = (\hbar/m\omega)^{1/2}$ is
much larger than the inter-particle distance, the density profile
in this inhomogeneous situation may be obtained from the Thomas-Fermi
approximation \cite{dunjko01}
\begin{equation}
  \mu\left[n(x)\right] + V(x) = \mu[n(0)]
  \label{TF}
\end{equation}
where $\mu[n]$ is the chemical potential of the homogeneous
system. The central density $n(0)$ and the associated radius
$R=\left(2\mu[n(0)]/m\omega^{2}\right)^{1/2}$ of the cloud is
obtained from the normalization condition
\begin{equation}
  N= \int_{-R}^{R} dx \: n(x) = 2 R \int_{0}^{n(0)} dn
  \sqrt{1-\frac{\mu[n]}{\mu[n(0)]}}
  \label{Nvsmu}
\end{equation}
with $N$ the particle number, provided the relation $\mu[n]$ is
known explicitly.

In the absence of an optical lattice the density profiles are
smooth functions of the coupling $\gamma$ \cite{dunjko01}. In the
limit $\gamma\gg1$, $\mu[n] \rightarrow \mu_{\rm\scriptscriptstyle
F} [n]=(\hbar \pi n)^{2}/2m$ approaches the Fermi energy of an
ideal Fermi gas with density $n$, resulting in a profile
$n(x)=(2N/\pi R)\sqrt{1-(x/R)^{2}}$ with radius $R = (2 N)^{1/2}
l$. Adding now a periodic potential which is nearly
commensurate with the density in the trap center opens a gap
$\Delta\mu$ in the chemical potential. The density profile will
then exhibit a flat, incompressible regime in the trap center. In
order to find the detailed shape of the density profile, we
approximate the sine-Gordon model in its strong coupling, gapped
phase by a gas of noninteracting relativistic Fermions of mass
$M$; in the relevant regime $1<K<2$, this approximation is known
to work extremely well \cite{kehrein99_01}. The chemical potential as
a function of density can then be calculated explicitly,
\begin{equation}
   \mu[n]\approx \mu_{\rm\scriptscriptstyle F}[n]
   +\frac{\Delta\mu}{2}
   f\biggl(\frac{4KE_{r}}{\Delta\mu}\frac{\delta n}{n_c}\biggr)
   \label{chempot}
\end{equation}
with $\delta n=n-n_{c}$ ($n_{c} = 2/\lambda$). The dimensionless
function
\begin{equation}
   f(z)=\pm\left( 1+z^{2}\right)^{1/2}-z
\end{equation}
is discontinous at $z=0$ and incorporates, via (\ref{chempot}),
the jump $\Delta\mu$ in the chemical potential at $n_{c}$
associated with the incompressible commensurate
state. Using this approximation, the density profile and the
fraction of particles participating in the commensurate phase
derive from an integration of (\ref{Nvsmu}) with (\ref{chempot}),
see Fig.~\ref{excitationgap}(b). 
Knowledge of the locked fraction
$\Delta N/N$ plays an important role in the experimental detection
of a commensurate Mott phase. This can be achieved by measuring
the excitation gap through a phase gradient method as done
previously for the Bose-Hubbard transition \cite{greiner02}.
Alternatively, it should be possible to directly observe the
increase in the long-range translational order in the Mott phase
via Bragg diffraction \cite{birkl95,weidemuller95}; 
in either case the fraction $\Delta
N/N$ determines the experimentally available signal. The latter
can be further enhanced by generating an array of parallel `atom
wires' with the help of a strong 2D optical transverse lattice.
Using numbers similar to those in the recent experiment by Greiner
{\it et al.} \cite{greiner02}, it is possible to generate several
thousand parallel 1D wires with a transverse confining frequency
$\nu_{\perp}=20$ kHz. A longitudinal harmonic trap with 
frequency $\nu=40$ Hz then
encloses $N \approx 50$ atoms in each 1D wire; the associated
central density in the absence of a longitudinal periodic
potential is $n(0)=2~\mu\hbox{m}^{-1}$ for $\gamma \gg 1$,
commensurate with the lattice constant $\lambda/2 \approx 0.5~\mu$m
of a typical optical lattice \cite{greiner02}. A weak periodic
potential will then lead to an incompressible Mott state in the
center of the cloud, provided the parameter $\gamma={2a_{s}}
/{n(0)l_{\perp}^{2}}$ is larger than the critical value
$\gamma_{c} \approx 3.5$. For $^{87}$Rb with a scattering 
length $a_{s}\approx
5$ nm, the resulting $\gamma$ is equal to one, i.e., not quite in
the required range. As noted already by Petrov {\it et al.}
\cite{petrov00}, however, larger and in particular tunable values
of $\gamma$ may be realized by changing $a_{s}$ via a Feshbach
resonance as present, e.g., in $^{85}$Rb.

In conclusion we have shown that a commensurate Mott state can be
realized in dilute 1D BEC's already with an arbitrary weak lattice
potential, provided that the ratio $\gamma$ between the
interaction and kinetic energies is larger than a critical value
$\gamma_{c}\approx 3.5$. This instability  provides a new and
experimentally accessible tool for the quantitative
characterization of 1D atomic gases in the strongly correlated
`Tonks gas' limit. Also, the observation of a Mott state in a
regime where the atoms are not confined to discrete lattice sites
would give direct evidence for the granularity of matter in strongly
interacting dilute gases.

It is a pleasure to acknowledge fruitful discussions with I.\
Bloch, T.\ Esslinger, M.\ Greiner and S.\ Kehrein. This work was
supported by the DFG Schwerpunkt {\it Ultrakalte Quantengase}.

 %\bibliographystyle{/home/buechler/Refbib/apsrev}
 %\bibliography{/home/buechler/Refbib/journals,/home/buechler/Refbib/ref}

%\end{multicols}
\end{document}